\documentstyle[prd,aps,psfig]{revtex}
\def\be{\begin{equation}}
\def\ee{\end{equation}}
\def\ba{\begin{eqnarray}}
\def\ea{\end{eqnarray}}
\newcommand{\lets}{\raise.3ex\hbox{$<$\kern-.75em\lower1ex\hbox{$\sim$}}}
\begin{document}
%
%
\draft
\title{Qualitative Signals of New Physics in $B - \overline{B}$ Mixing}
\author{Jo\~ao P.\ Silva\footnote{Permanent address: 
	Instituto Superior de Engenharia de Lisboa,
	Rua Conselheiro Em\'{\i}dio Navarro,
	1900 Lisboa, Portugal.}}
\address{Stanford Linear Accelerator Center, Stanford University, Stanford,
	 California 94309}
\author{Lincoln Wolfenstein}
\address{Department of Physics,
Carnegie Mellon University, Pittsburgh, Pennsylvania 15213}
\date{\today}
\maketitle
\begin{abstract}
It is expected that,
within the next three years,
the determination of the CKM matrix elements with the least theoretical
uncertainty will arise from the measurements of CP violation
in $B_d \rightarrow J/\psi K_S$ decays and from
$B_s - \overline{B_s}$ mixing.
If there is significant new physics in $B - \overline{B}$ mixing,
then those experiments will not yield the correct values for the
CKM matrix elements.
As a result,
a qualitative signal of new physics may appear in the CP violation of
decays like $B_d \rightarrow \pi^+ \pi^-$.
\end{abstract}


\section{Introduction}
\label{sec:intro}

The strongest constraints on the CKM matrix likely to be
achieved in the next three years will arise from the values
of $\sin{2 \beta}$ and $x_s$.
Results from CDF, Babar and
Belle should provide $\sin(2 \beta)$ with an error of
$0.08$ or less \cite{BaBarbook}.
If $x_s$ is not too large,
then it is hoped that $x_s$ will be measured to better than
$10\%$ in Run 2 of CDF \cite{CDF}.
For the analysis of the CKM matrix the useful quantity is
\be
\frac{x_s}{x_d} = \left( \frac{V_{ts}}{V_{td}} \right)^2 K,
\ee
where $K$ is a SU(3) correction factor,
calculated to be $1.3$ with a claim by lattice calculations of a
$7\%$ error \cite{Mackenzie}.

In terms of CKM parameters \cite{Wolfpara},
\be
\sin{2 \beta} = \frac{2\, \eta\, (1-\rho)}{(1-\rho)^2+\eta^2},
\label{sin2beta}
\ee
and
\be
\frac{x_s}{x_d} = \frac{K}{\lambda^2 R_t^2}
\label{xsxd}
\ee
where
\be
R_t = \sqrt{(1-\rho)^2 + \eta^2}
= \frac{\sin{\gamma}}{\sin{(\beta+\gamma)}}.
\ee
These two results will produce a small allowed region for the
CKM parameters,
assuming that they are consistent with the present loose 
constraints from $|V_{ub}|$ and $\epsilon_K$.

It is interesting to note that both Eqs.~(\ref{sin2beta})
and (\ref{xsxd}) have to do with $B - \overline{B}$
mixing.
Thus,
if there is a significant new physics contribution to 
$B - \overline{B}$ mixing,
then these measurements provide {\em no constraint on CKM parameters}.
Rather,
they would provide wrong values for the parameters,
which we denote by
$\tilde{\rho}$,
$\tilde{\eta}$,
$\tilde{\beta}$,
$\tilde{\gamma}$,
and
$\tilde{R}_t$.
Naturally,
\be
\tilde{R}_t = \sqrt{(1-\tilde{\rho})^2 + \tilde{\eta}^2}
= \frac{\sin{\tilde{\gamma}}}{
\sin{(\tilde{\beta}+\tilde{\gamma})}}.
\ee
We also define
\be
\tilde{R}_b = \sqrt{{\tilde{\rho}}^2 + {\tilde{\eta}}^2}
= \frac{\sin{\tilde{\beta}}}{\sin{(\tilde{\beta}+\tilde{\gamma})}}.
\ee

A new physics contribution to $B_d - \overline{B_d}$ mixing,
described by an effective superweak-like interaction
\be
G_{bd}\ \bar{b} {\cal O} d\ \bar{b} {\cal O} d\ + \mbox{h.c.},
\label{superweak}
\ee 
with $G_{bd}$ of order $10^{-7} G_F$ to $10^{-8} G_F$,
will affect both Eqs.~(\ref{sin2beta}) and (\ref{xsxd}).
Eq.~(\ref{sin2beta}) is also sensitive to new physics in
$B_s - \overline{B_s}$ mixing.
Considering only new physics of the form in Eq.~(\ref{superweak}),
the measurements of $\sin{2 \beta}$ and $x_s$ really
determine the $B_d - \overline{B_d}$ matrix $M_{12}$:
\ba
\mbox{Im} M_{12} & \propto & 2\, \tilde{\eta}\, (1 - \tilde{\rho}),
\label{eqa}
\\
\left| M_{12} \right| & \propto & (1 - \tilde{\rho})^2 + \tilde{\eta}^2,
\label{eqb}
\\
\mbox{Re} M_{12} & \propto &
\pm \left[ (1 - \tilde{\rho})^2 - \tilde{\eta}^2 \right].
\label{eqc}
\ea
The CKM fit requires the positive value; the negative value corresponds
to the well known ambiguity
$\tilde{\beta} \rightarrow \pi/2 - \tilde{\beta}$
\cite{ambiguity}.

The present limit on $x_s$ yields the result that
$\tilde{R}_t\, \raise.3ex\hbox{$<$\kern-.75em\lower1ex\hbox{$\sim$}}\, 1$.
If we assume that $x_s$ will be measured to be between $20$ and $30$,
then $\tilde{R}_t$ will lie between $1.0$ and $0.8$,
corresponding to $\tilde{\gamma}$ in the region between
$80^\circ$ and $50^\circ$.
On the other hand,
it has been suggested on the basis of the measured
$B$ decay rates that $\gamma$ is greater than $90^\circ$ \cite{largegamma}.
If this were true,
then it could be a sign of new physics.

As a possible way of finding a sign of this new physics,
we consider the asymmetry in the time dependent decays
$B_d (\overline{B_d}) \rightarrow \pi^+ \pi^-$.
In Fig.~1 the solid line shows the expected asymmetry for
values of $\tilde{R_t}$ between $1.0$ and $0.8$,
with $\tilde{\beta} = 25^\circ$
(this corresponds to values for $\tilde{R_b}$ which lie between
$0.422$ and $0.436$). 
The values of the asymmetry are well above zero.
\begin{figure}
\centerline{\psfig{figure=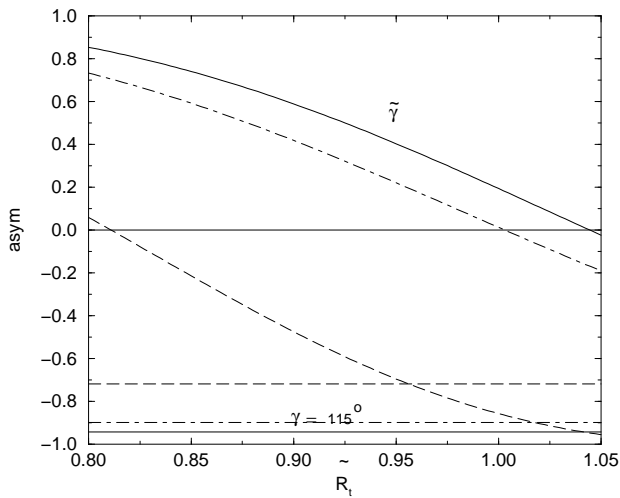,height=3in}}
\caption{The figure shows the interference CP-asymmetry
in $B_d (\overline{B_d}) \rightarrow \pi^+ \pi^-$ decays.
The curves correspond to the asymmetry obtained with $\tilde{\gamma}$
for strong phases $\Delta = 0^\circ$ (solid line),
$\Delta = 45^\circ$ (dot-dashed line),
and $\Delta = 180^\circ$ (dashed line).
For comparison,
we also show the horizontal lines corresponding to the values of
the asymmetry obtained with $\gamma = 115^\circ$,
for the same three values of $\Delta$.
\label{fig:1}}
\end{figure}
In contrast,
if the true $\gamma \sim 115^\circ$,
then the expected asymmetry is approximately $-0.9$.
Thus,
a discovery of such a large negative asymmetry in this scenario could be
a signal of new physics in $B - \overline{B}$ mixing.
Note that the penguin contribution plays an important role in this analysis.
In the absence of penguins the first asymmetry would be
$\sin{2(\tilde{\beta}+\tilde{\gamma})}$,
which is already significantly larger than
$\sin{2(\tilde{\beta}+\gamma)}$.
However,
inclusion of the penguin separates the asymmetries further
by increasing the first asymmetry well above 
$\sin{2(\tilde{\beta}+\tilde{\gamma})}$.
This makes the difference even more striking.

Details of this calculation are given in the appendix.
We have used a method proposed by Silva and Wolfenstein \cite{SW94}
in which one invokes SU(3) to relate the tree and penguin contributions
in the $B_d \rightarrow \pi^+ \pi^-$ decays to those in the
$B_d \rightarrow K^\pm \pi^\mp$ decays.
The details of this calculation depend on SU(3),
factorization and were originally performed assuming
that the strong phase shift between the penguin and the tree contributions 
in the $B_d \rightarrow \pi^+ \pi^-$ and in the
$B_d \rightarrow K^\pm \pi^\mp$ channels is equal to a common
value $\Delta$ and, moreover,
that $\Delta = 0^\circ$.
Recently,
this analysis has been revisited by Fleischer \cite{Flei},
allowing $\Delta$ to take any value.
Since we are interested in a large qualitative effect,
small changes in the assumptions made in the calculation will not affect
the conclusion.
For example, 
setting $\Delta = 45^\circ$ yields the dash-dotted curve in Fig.~1.
There is, however,
one critical assumption: the sign of the penguin term relative
to that of the tree.
If we choose the negative sign,
or equivalently let $\Delta = 180^\circ$,
we obtain the dashed curve in Fig.~1.
Assuming this sign,
the asymmetry for $\gamma=115^\circ$ is about $-0.7$.
Thus,
except for low values of $\tilde{R}_t$ (large values of $x_s$),
the large qualitative difference disappears.

As discussed by Fleischer \cite{Flei},
allowing for any value of $\Delta$,
the measured asymmetry allows a limited range of values for $\gamma$.
This leads us to consider the following three possibilities:
\begin{itemize}
\item If a sizeable positive asymmetry is found,
this would be consistent with values of $\tilde{\gamma}$
well below $90^\circ$,
as postulated in our scenario,
and rule out values greater than $90^\circ$.
\item A very large negative asymmetry would be consistent with
a true value of $\gamma$ in the neighborhood of $115^\circ$,
as has been proposed \cite{largegamma},
and could rule out values of $\tilde{\gamma}$ corresponding
to $\tilde{R}_t < 0.95$.
In this case we could have a signal of new physics independent 
of any assumption about $\Delta$.
\item A negative asymmetry in the neighborhood of $-0.5$ could
be interpreted in different ways: (1) if one believes from
factorization that $\cos{\Delta}$ is close to $+1$,
then one obtains $\gamma$ slightly above $90^\circ$
and this could be a sign of new physics in our scenario;
(2) the true value of $\gamma$ is really $\tilde{\gamma}$
for values of $\tilde{R}_t$ above $0.85$ but $\cos{\Delta}$
is close to $-1$.
\end{itemize}
It is interesting to note that,
if $\cos{\Delta}$ is close to $-1$,
then the motivation for a value $\gamma > 90^\circ$ given
in Ref.~\cite{largegamma} may disappear.

As a particular example,
let us take $\tilde{\eta} = 0.4$ and $(1-\tilde{\rho})=0.8$,
corresponding to $\tilde{R}_t = 0.89$,
$\tilde{\gamma}=63.5^\circ$, and $\tilde{\beta} = 26.5^\circ$.
If the true values were  $\eta = 0.4$ and $(1-\rho)=1.2$,
$\beta=18.5^\circ$, and $\gamma = 116.5^\circ$,
then,
in order to fit the $B_d - \overline{B_d}$ mixing from
Eqs.~(\ref{eqa})--(\ref{eqc}),
with the positive sign in Eq.~(\ref{eqc}),
we would need
\ba
\frac{\left(\mbox{Im} M_{12}\right)_{sw}}{
\left(\mbox{Im} M_{12}\right)_{\rm CKM}}
&=&
- \frac{1}{3}
\nonumber\\
\frac{\left(\mbox{Re} M_{12}\right)_{sw}}{
\left(\mbox{Re} M_{12}\right)_{\rm CKM}}
&=&
- \frac{5}{8},
\ea
where the subscript $sw$ denotes a superweak-like contribution,
as in Eq.~(\ref{superweak}),
and the subscript ${\rm CKM}$ denotes the Standard Model contribution.

Another possible interpretation of the large negative asymmetry
in this scenario would be that the true value
of $\mbox{Re} M_{12}$ is given by the minus sign in Eq.~(\ref{eqc}).
In that case,
the phase $2 \tilde{\beta}$ would be about $130^\circ$ instead
of $50^\circ$.
Then the true value of $\gamma$ would have to be around $70^\circ$
in order to explain the large negative asymmetry.
Therefore,
the main new contribution would be
\be
\left(\mbox{Re} M_{12}\right)_{sw} \sim - 
2 \left(\mbox{Re} M_{12}\right)_{\rm CKM}.
\ee

\section{Conclusions}
\label{sec:conclusions}

In the Standard Model the best determination of the
CKM matrix elements,
from $\sin{2 \beta}$ and $x_s$,
depends on the absence of new physics effects in the mixing.
Therefore,
it is possible that experiments that are really sensitive to the phases
of decay amplitudes may show qualitative deviations from expectations
based on  $\sin{2 \beta}$ and $x_s$.
One example is discussed in this paper,
where we have stressed the usefulness of a {\em qualitative}
measurement of the CP-violating asymmetry in 
$B_d \rightarrow \pi^+ \pi^-$ decays.

\acknowledgments

This work is supported in part by the Department of Energy 
under contracts DE-AC03-76SF00515 and DE-FG02-91-ER-40682.
The work of J.\ P.\ S.\ is supported in part by Fulbright,
Instituto Cam\~oes, and by the Portuguese FCT, under grant
PRAXIS XXI/BPD/20129/99	and contract CERN/S/FIS/1214/98.
L.\ W.\ thanks SLAC for support during the time this work was
carried out.

\appendix

%

\section*{Determining the asymmetry in
$B_d \rightarrow \pi^+ \pi^-$ decays}
\label{app:A}

For completeness,
we include in this section a description of the
method proposed by Silva and Wolfenstein \cite{SW94}
to estimate the asymmetry in the time dependent decays
$B_d (\overline{B_d}) \rightarrow \pi^+ \pi^-$.
In this method one uses two observables as input:
$\sin{2 \tilde{\beta}}$ obtained from the CP-violating 
asymmetry in $B_d \rightarrow J/\psi K_S$ decays;
and 
\be
R \equiv \frac{\Gamma[B_d \rightarrow K^+ \pi^-] + 
\Gamma[\overline{B_d} \rightarrow K^- \pi^+]}{
\Gamma[B_d \rightarrow \pi^+ \pi^-] +
\Gamma[\overline{B_d} \rightarrow \pi^+ \pi^-]},
\label{definition-of-R-SW}
\ee
recently measured by CLEO to be $R = 4.0 \pm 2.2$ \cite{CLEO},
where we have added the errors in quadrature.
In addition,
one uses the spectator approximation and SU(3) 
(in fact a U-spin rotation interchanging the $d$ and $s$ quarks)
to relate the tree dominated process $B_d \rightarrow \pi^+ \pi^-$
to the penguin dominated $B_d \rightarrow K^\pm \pi^\mp$.
Here we will follow the notation of Ref.~\cite{BLS}.

The relevant decay amplitudes may be written as
\ba
A(B_d \rightarrow \pi^+ \pi^-)
& = &
\frac{V_{ub}^\ast V_{ud}}{|V_{ub} V_{ud}|} T +
\frac{V_{tb}^\ast V_{td}}{|V_{tb} V_{td}|} P e^{i \Delta}
=
e^{i \gamma} T \left[ 1 + r e^{i(-\beta - \gamma + \Delta)}\right],
\label{Bpipi-amp}
\\
A(B_d \rightarrow K^+ \pi^-)
& = &
\frac{V_{ub}^\ast V_{us}}{|V_{ub} V_{us}|} T^\prime +
\frac{V_{tb}^\ast V_{ts}}{|V_{tb} V_{ts}|} P^\prime e^{i \Delta^\prime}
=
e^{i \gamma} T^\prime \left[ 1 - r^\prime 
e^{i(- \gamma + \Delta^\prime)}\right].
\label{BKpi-amp}
\ea
Here $\Delta$ and $\Delta^\prime$ are strong phases,
while $r = P/T$ and $r^\prime = P^\prime/T^\prime$
are the ratios of the magnitudes of the penguin to the tree
contributions in each channel, respectively.
Therefore \cite{BLS},
\be
R =
\frac{{T^\prime}^2}{T^2}\,
\frac{{r^\prime}^2 - 2 r^\prime \cos{\gamma} \cos{\Delta^\prime} + 1}{
1 + 2 r \cos{\left( \beta + \gamma \right)} \cos{\Delta} + r^2}.
\ee

Using U-spin we may relate the two decay channels through
\ba
\Delta^\prime & = & \Delta,
\nonumber\\
\frac{T^\prime}{T} & = & 
\left| \frac{V_{us}}{V_{ud}} \right| \frac{f_K}{f_\pi} \approx
\lambda \frac{f_K}{f_\pi},
\\
\frac{P^\prime}{P} & = &
\left| \frac{V_{ts}}{V_{td}} \right| \frac{f_K}{f_\pi}
\approx
\frac{1}{\lambda R_t}
\frac{f_K}{f_\pi}.
\label{Pprime/P}
\ea
In the last two expressions we have used factorization 
to estimate the ratio of matrix elements in the two channels.
This provides a first order estimate of the SU(3) breaking effects.
Therefore,
defining
\be
a \equiv
\frac{r^\prime}{r} = \frac{1}{\lambda^2 R_t},
\ee
we obtain
\be
R = \left( \lambda \frac{f_K}{f_\pi}\right)^2
\frac{a^2 r^2 - 2\, a\, r \cos{\gamma} \cos{\Delta} + 1}{
1 + 2 r \cos{(\beta + \gamma)} \cos{\Delta} + r^2}.
\label{masterR}
\ee
Here,
we have used the true values of $\beta$ and $\gamma$ ($R_t$)
as they appear in the CKM matrix.
Eq.~(\ref{masterR}) determines $r$.
Finally,
the time dependent asymmetry in 
$B_d (\overline{B_d}) \rightarrow \pi^+ \pi^-$
decays is given by
\be
{\rm asym} = \frac{2\, \mbox{Im} \lambda_f}{1 + |\lambda_f|^2}
\label{masterasym}
\ee
where
\be
\lambda_f = \frac{q}{p}
\frac{A(\overline{B_d} \rightarrow \pi^+ \pi^-)}{
A(B_d \rightarrow \pi^+ \pi^-)}
=
- e^{-2 i(\tilde{\beta}+\gamma)}
\frac{1 + r e^{i(\beta+\gamma+\Delta)}}{
1 + r e^{i(-\beta-\gamma+\Delta)}}.
\label{masterlambdaf}
\ee
Notice that we have used $q/p = - \exp{(- 2 i \tilde{\beta})}$ 
(as opposed to the true CKM phase $\beta$)
because any new physics appearing in the mixing will affect
the interference CP-violating asymmetries in both the
$B_d \rightarrow \pi^+ \pi^-$ and 
$B_d \rightarrow J/\psi K_S$ channels,
through the same $q/p$.
In the presence of new physics $\beta$ in Eqs.~(\ref{Bpipi-amp})
and (\ref{BKpi-amp}) need not be the same as $\tilde{\beta}$.

Fig.~1 was obtained in the following way. We have set
$\tilde{\beta}=25^\circ$.
Then, for each value of $\tilde{R}_t$ between
$0.8$ and $1$,
we calculate $\tilde{\gamma}$ and $\tilde{R}_b$.
Our choice of $\tilde{\beta}$ guarantees that
$\tilde{R}_b$ lies within the range currently allowed by the
measurement of $|V_{ub}|$.
We have taken $R=3$,
which is in the lower side of the range allowed by CLEO \cite{CLEO},
in order to facilitate comparison with Ref.~\cite{SW94}.
Next,
we consider the scenario in which the true value of $\gamma$
is really $115^\circ$.
The corresponding value of $\beta$ is obtained by requiring
$R_b=\tilde{R_b}$,
thus continuing to satisfy the $|V_{ub}|$ measurement.
For each case,
we can determine $r$ from Eq.~(\ref{masterR}),
$\lambda_f$ from Eq.~(\ref{masterasym}),
and the asymmetry from Eq.~(\ref{masterlambdaf}).



\begin{references}
%
\bibitem{BaBarbook}
See, for instances,
BABAR Collaboration,
{\it The BaBar physics book},
edited by P.\ F.\ Harrison and H.\ R.\ Quinn
(SLAC, Stanford, 1998).
%
\bibitem{CDF}
J.\ Kroll,
talk given at the {\it $B$ Physics at Tevatron Workshop},
September 1999.
%
\bibitem{Mackenzie}
P.\ Mackenzie,
talk given at the {\it $B$ Physics at Tevatron Workshop},
September 1999.
%
\bibitem{Wolfpara}
L.\ Wolfenstein,
Phys.\ Rev.\ Lett. {\bf 51}, 1945 (1983).
%
\bibitem{ambiguity}
L.\ Wolfenstein,
in {\it Honolulu 1997: B physics and CP violation},
pp.\ 476-485.
Y.\ Grossman, Y.\ Nir, and M.\ P.\ Worah,
Phys.\ Lett.\ B {\bf 407}, 307 (1997).
Y.\ Grossman and H.\ R.\ Quinn,
Phys.\ Rev.\ D {\bf 56}, 7259 (1997).
L.\ Wolfenstein,
Phys.\ Rev.\ D {\bf 57} 6857 (1998). 
B.\ Kayser and D.\ London,
hep-ph/9909560,
to appear in Phys.\ Rev.\ D.
For a review also \cite{BLS}.
%
\bibitem{largegamma}
N.\ G.\ Deshpande, X.\ G.\ He, W.\ S.\ Hou, and S.\ Pakvasa,
Phys.\ Rev.\ Lett. {\bf 82}, 2240 (1999).
X.\ G.\ He, W.\ S.\ Hou, and K.\ C.\ Yang,
Phys.\ Rev.\ Lett. {\bf 83}, 1100 (1999).
W.\ S.\ Hou, J.\ G.\ Smith, and F.\ W\"{u}rthwein,
National Taiwan University preprint number NTUHEP-99-25,
hep-ex/9910014.
%
\bibitem{SW94}
J.\ P.\ Silva and L.\ Wolfenstein,
Phys.\ Rev.\ D {\bf 49}, R1151 (1994).
%
\bibitem{Flei}
R.\ Fleischer,
DESY preprint number DESY 00-014,
hep-ph/0001253.
%
\bibitem{CLEO}
CLEO Collaboration,
D.\ Cronin-Hannessy {\it et al.},
CLEO preprint number CLEO 99-18,
hep-ex/0001010.
%
\bibitem{BLS}
G.\ C.\ Branco, L.\ Lavoura, and J.\ P.\ Silva,
{\it CP Violation}\, (Oxford University Press, Oxford, 1999).
%
\end{references}
\end{document}